\documentclass[aps,twocolumn,pra,showpacs]{revtex4}
\usepackage{bbm}
\usepackage{amsmath}
\usepackage{graphics}
\usepackage{epsfig}
\def\RR{\ensuremath{\mathbbm{R}}}
\def\CC{\ensuremath{\mathbbm{C}}}
\def\id{\ensuremath{\mathbbm{1}}}
\def\LZR{\ensuremath{\mathrm{L}^{\!\rule[-0.5ex]{0mm}{0mm}2}(\RR)}}
\def\cE{{\cal E}}
\def\cH{{\cal H}}
\def\cC{{\cal C}}
\def\cP{{\cal P}}
\def\cS{{\cal S}}
\def\cG{{\cal G}}
\def\qed{\rule{1ex}{1ex}}
\def\tr{\mathrm{tr}}

\def\del{\partial}

\def\equivv{\Leftrightarrow}
\renewcommand{\Re}{\mathrm{Re}}
\renewcommand{\Im}{\mathrm{Im}}
\def\Proof{\textsc{Proof: }}
\newtheorem{theorem}{Theorem}
\newtheorem{lemma}{Lemma}
\newtheorem{definition}{Definition}

\newtheorem{observation}{Observation}

\newcommand{\Eqref}[1]{Eq.~(\ref{#1})}
\newcommand{\Ineqref}[1]{Ineq.~(\ref{#1})}
\newcommand{\Eqsref}[1]{Eqs.~(\ref{#1})}
\newcommand{\Ineqsref}[1]{Ineqs.~(\ref{#1})}
\newcommand{\Thref}[1]{Theorem~\ref{#1}}
\newcommand{\Leref}[1]{Lemma~\ref{#1}}

\newcommand{\Figref}[1]{Fig.~\ref{#1}}
\begin{document}
\title{Separability Properties of Three-mode Gaussian States}

\author{G. Giedke$^{(1)}$, B. Kraus$^{(1)}$, M. Lewenstein$^{(2)}$,
and J. I. Cirac$^{(1)}$}

\affiliation{(1) Institut f\"ur Theoretische Physik, Universit\"at
Innsbruck, A-6020 Innsbruck, Austria\\
(2) Institut f\"ur Theoretische Physik, Universit\"at
Hannover, 30163 Hannover, Germany}

\begin{abstract}
We derive a necessary and sufficient condition for the separability of
tripartite three mode Gaussian states, that is easy to check for any
such state. We give a classification of the separability properties
of those systems and show how to determine for any state to which
class it belongs. We show that there exist genuinely tripartite bound
entangled states and point out how to construct and prepare such
states.
\end{abstract}


\maketitle

\section{Introduction}
Entanglement of composite quantum systems is central to both the
peculiarities and promises of quantum information. Consequently, the
study of entanglement of bi- and multipartite systems has been the
focus of research in quantum information theory. While pure state
entanglement is fairly well understood, there are still many open
questions related to the general case of mixed states. The furthest
progress has been made in the study of systems of two qubits: it has
been shown that a state of two qubits is separable if and only if its
partial transpose is positive (PPT-property) \cite{QBSepCrit} and a
closed expression for the entanglement of formation was derived
\cite{Wootters}. Moreover, it was shown \cite{QBdist} that all entangled
states of two qubits can be \emph{distilled} into maximally entangled pure
states by local operations. This property of distillability is of great
practical importance, since only the distillable states are useful for
certain applications such as long-distance quantum communication, quantum
teleportation or cryptography \cite{applics}.

In higher dimensions much less is known: the PPT-property is no longer
sufficient for separability as proved by the existence of PPT
entangled states in $\CC^2\otimes\CC^4$ systems \cite{pptes}. These
states were later shown to be \emph{bound entangled} \cite{BE}: even
if two parties (Alice and Bob) share an arbitrarily large supply of
such states, they cannot transform (``distill'') it into even a single
pure entangled state by local quantum operations and classical
communication. Meanwhile, a number of additional necessary or
sufficient conditions for inseparability have been found for finite
dimensional bipartite systems, which use properties of the range and
kernel of the density matrix $\rho$ and its partial transpose
$\rho^{T_A}$ to establish separability (\cite{primer} and references
therein).

When going from two to more parties, current knowledge is even more
limited.  Pure multipartite entanglement was first considered in
\cite{GHZ}. A classification of $N$-partite mixed states according to
their separability properties has been given \cite{classent}.  But
even for three qubits there is currently no general way to decide to
which of these classes a given state belongs \cite{Acin01}. Results on
bound entanglement \cite{upb} and entanglement distillation
\cite{multipartact} for multi-party systems have been obtained.

Recently increasing attention was paid to infinite dimensional
systems, the so-called continuous quantum variables (CV), in
particular since the experimental realization of CV quantum
teleportation \cite{CVQTelTh,CVQTelExp}. Quantum information with CV
in general is mainly concerned with the family of \emph{Gaussian
states}, since these comprise essentially all the experimentally
realizable CV states. A practical advantage of CV systems is the
relative ease with which entangled states can be generated in the lab
\cite{CVQTelExp,Leuchs}. First results on separability and
distillability of Gaussian states were reported in
\cite{Reid,Duan99,Sim99,DCVS,GBE,compEM,GSepCrit}. One finds striking
similarities between the situations of two qubits and two one--mode CV
systems in a Gaussian state: PPT is necessary and sufficient for
separability \cite{Duan99,Sim99}, and all inseparable states are
distillable \cite{DCVS}. Generalizing the methods reviewed in
\cite{primer} it was shown that for more than two modes at either side
PPT entangled states exist \cite{GBE}. In \cite{compEM} a computable
measure of entanglement for bipartite Gaussian states was derived.

The study of CV multipartite entanglement was initiated in
\cite{CVmultipent,CVghz}, where a scheme was suggested to create pure CV
$N$-party entanglement using squeezed light and $N-1$ beamsplitters.
In fact, this discussion indicates that tripartite entanglement has
already been created (though not investigated or detected) in the CV
quantum teleportation experiment \cite{CVQTelExp}.

In this paper we provide a complete classification of tri-mode
entanglement (according to the scheme \cite{classent}) and obtain --
in contrast to the finite dimensional case -- a simple, directly
computable criterion that allows to determine to which class a given
state belongs. We show that none of these classes are empty and in
particular provide examples of genuine tripartite bound entangled
states, i.e. states of three modes $A$, $B$, and $C$ that are
separable whenever two parties are grouped together but cannot be
written as a mixture of tripartite product states.

Before we can derive our results we need to introduce some notation
and collect a number of useful facts about our main object of study:
Gaussian states.

\section{Gaussian States}\label{GaussianSec}

In quantum optics and in other scenarios described by continuous
quantum variables, not all states on the infinite dimensional Hilbert
space are equally accessible in current experiments. In fact, the set
of \emph{Gaussian states} comprises essentially all genuinely CV
states that can currently be prepared in the lab. This, and the
mathematical simplicity of these states are the reasons why CV quantum
information has so far considered almost exclusively Gaussian states,
as will the present paper. This section summarizes results on Gaussian
states that we need in the following and introduces some notation.

We consider systems composed of $n$ distinguishable infinite
dimensional subsystems, each with Hilbert space $\cH = \LZR$. These
could be implemented quantum optically by different modes of the
electromagnetic field, hence each of these subsystems will be referred
to as a ``mode''. To each mode belong the two canonical observables
$X_k, P_k, k=1,\dots,n$ with commutation relation
$[X_k,P_k]=i$. Defining $R_k=X_k, R_{n+k}=P_k$ these relation are
summarized as $[R_k,R_l]=-iJ_{kl}$, using the antisymmetric $2n\times
2n$ matrix
\begin{equation}\label{J}
J = \left( \begin{array}{cc}
0&-\id\\
\id&0\end{array} \right).
\end{equation}
which plays an important role in the following calculations
\cite{fndimofJ}.

For such systems, it is convenient to describe the
state $\rho$ by its characteristic function
\begin{equation}
\chi(x) = {\rm tr} [\rho D(x)].
\end{equation}
Here $x = (q,p),\, q,p\in\RR^n$ is a real vector, and
\begin{equation}
D(x)=e^{-i\sum_{k} (q_k X_k + p_k P_k)}.
\end{equation}
The characteristic function contains all the information about the
state of the system, that is, one can construct $\rho$ knowing
$\chi$. Gaussian states are exactly those for which $\chi$ is a
Gaussian function of the phase space coordinates $x$ \cite{Manu},
\begin{equation}\label{charfct}
\chi(x) = e^{-\frac{1}{4}x^T \gamma x - id^Tx},
\end{equation}
where $\gamma$ is a real, symmetric, strictly positive matrix, the
\emph{correlation matrix} (CM), and $d\in\RR^{2n}$ is a real vector, the
\emph{displacement}. Note that both $\gamma$ and $d$ are directly
measurable quantities, their elements $\gamma_{kl}$ and $d_k$ are
related to the expectation values and variances of the operators
$R_k$. A Gaussian state is completely determined by $\gamma$ and
$d$. Note that the displacement of a (known) state can always be
adjusted to $d=0$ by a sequence of unitaries applied to individual
modes. This implies that $d$ is irrelevant for the study of nonlocal
properties. Therefore we will occasionally say, e.g., that ``a CM is
separable'' when the Gaussian state with this CM is separable.  Also,
from now on in this paper ``state'' will always mean ``Gaussian
state'' (unless stated otherwise).

Not all real, symmetric, positive matrices $\gamma$ correspond to the
CM of a physical state. There are a number of equivalent ways to
characterize physical CMs, which will all be useful in the
following. We collect them in

\begin{lemma} (Correlation Matrices)\\
\label{physical} For a real, symmetric $2n\times2n$ matrix $\gamma>0$ the
following statements are equivalent:
\begin{subequations}\label{physicalconds}
\begin{equation}\label{i}
\mbox{$\gamma$ is the CM of a physical state},
\end{equation}
\begin{equation}\label{ii}
\gamma+J\gamma^{-1}J \geq 0,
\end{equation}
\begin{equation}\label{iii}
\gamma-i J \geq 0,
\end{equation}
\begin{equation}\label{iv}
\gamma = S^T(D\oplus D) S,
\end{equation}
for $S$ symplectic \cite{fnsympl} and $D\geq\id$ diagonal \cite{fnblock}.
\end{subequations}
\end{lemma}
\Proof\ $(\ref{i})\equivv(\ref{ii})$ see \cite{Manu};
$(\ref{i})\equivv(\ref{iii})$ see \cite{GBE};
$(\ref{i})\equivv(\ref{iv})$ see \cite[Prop. 4.22]{Folland}.\\

A CM corresponds to a pure state if and only if (iff) $D=\id$, i.e.\
iff $\det\gamma=1$ (e.g.\ \cite{Manu}). It is easy to see from
(\ref{iv}) that for pure states \Ineqref{ii} becomes an equality and
dim$\left[\ker(\gamma-iJ)\right]=n$. It is clear from \Eqref{iv} that
for every CM $\gamma$ there exists a pure CM $\gamma_0$ such that
$\gamma_0\leq\gamma$. This will allow us to restrict many proofs to
pure CMs only. Note that for a pure $2n\times2n$ CM $\gamma$ it holds
that $\tr
\gamma\geq 2n$.

A very important transformation for the study of entanglement is
partial transposition \cite{QBSepCrit}. Transposition is an example of
a positive but not completely positive map and therefore may reveal
entanglement when applied to part of an entangled system. On phase
space transposition corresponds to the transformation that changes the
sign of all the $p$ coordinates $(q,p)\mapsto\Lambda(q,p)=(q,-p)$
\cite{Sim99} and leaves the $q$'s unchanged. For $\gamma$ and $d$ this
means $(\gamma,d) \mapsto (\Lambda\gamma\Lambda,\Lambda d)$.  Using
this, the NPT-criterion for inseparability \cite{QBSepCrit} translates
very nicely to Gaussian states.  Consider a bipartite system
consisting of $m$ modes on Alice's side and $n$ modes on Bob's
($m\times n$-system in the following). Let $\gamma$ be the CM of a
Gaussian $m\times n$-state and denote by $\Lambda_A=\Lambda\oplus\id$
the partial transposition in A's system only. Then we have the
following criterion for inseparability:
\begin{theorem} (NPT criterion)\\ \label{nptcrit}
Let $\gamma$ be the CM of a $1\times n$ system, then $\gamma$
corresponds to a inseparable state if and only if
$\Lambda_A\gamma\Lambda_A$ is not a physical CM, i.e.\ if and only if
\begin{equation}\label{npt}
 \Lambda_A\gamma\Lambda_A\not\geq iJ.
\end{equation}
We say that $\gamma$ ``is NPT'' if (\ref{npt}) holds.
\end{theorem}
\Proof\ See \cite{Sim99} for $N=1$ and \cite{GBE} for the general
case.\\

Occasionally it is convenient to apply the orthogonal operation
$\Lambda_A$ to the right hand side of \Ineqref{npt} and write $\tilde
J_A\equiv\Lambda_AJ\Lambda_A$.

For states of at least two modes at both sides Condition\ (\ref{npt})
is still sufficient for inseparability, but no longer necessary as
shown by Werner and Wolf, who have considered a family of $2\times2$
entangled states with positive partial transpose \cite{GBE}. In the
same paper, it was shown that

\begin{theorem} (Separability of Gaussian States)\\\label{WernersepTh}
A state with CM
$\gamma$ is separable iff
there exist CMs $\gamma_A,\gamma_B$
such that
\begin{equation}\label{Wernersep}
\gamma\geq\gamma_A\oplus\gamma_B.
\end{equation}
\end{theorem}

It is observed in \cite{GBE} that if \Ineqref{Wernersep} can be
fulfilled, then the state with CM $\gamma$ can be obtained by local
operations and classical communication from the product state with CM
$\gamma_p=\gamma_A\oplus\gamma_B$, namely by mixing the states
$(\gamma_p,d)$ with the $d$'s distributed according to the Gaussian
distribution $\propto\exp\left[-d^T(\gamma-\gamma_p)^{-1}d\right]$.

Note that while \Thref{WernersepTh} gives a necessary and sufficient
condition for separability, it is not a practical criterion, since to
use it, we have to prove the existence or non-existence of CMs
$\gamma_A,\gamma_B$. Instead, a criterion would allow to directly
calculate from $\gamma$ whether the corresponding state is separable
or not. \Thref{WernersepTh} and its extension to the 3-party situation
are the starting point for the derivation of such a criterion for the
case of three-mode three-party states in the following main section of
this paper.

\section{Tri-mode Entanglement}\label{triSec}

When systems that are composed of $N>2$ parties are considered, there
are many ``types'' of entanglement due to the many ways in which the
different subsystems may be entangled with each other. We will use the
scheme introduced in \cite{classent}, to classify three-mode
tripartite Gaussian states.  The important point is that from the
extension of \Thref{WernersepTh} we can derive a simple criterion that
allows to determine which class a given state belongs to. This is in
contrast to the situation for three qubits, where up until now no such
criterion is known. In particular, we show that none of these classes
are empty and we provide an example of a genuine tripartite bound
entangled state, i.e.\ a state of three modes $A$, $B$, and $C$ that
is separable whenever two parties are grouped together but cannot be
written as a mixture of tripartite product states and therefore cannot
be prepared by local operations and classical communication of three
separate parties.

\subsection{Classification}

The scheme of \cite{classent} considers all possible ways to group the
$N$ parties into $m\leq N$ subsets, which are then themselves
considered each as a single party. Now, it has to be determined
whether the resulting $m$-party state can be written as a mixture of
$m$-party product states.  The complete record of the $m$-separability
of all these states then characterizes the entanglement of the
$N$-party state.

For tripartite systems, we need to consider four cases, namely the
three bipartite cases in which $AB$, $AC$, or $BC$ are grouped
together, respectively, and the tripartite case in which all $A$, $B$,
and $C$ are separate.  We formulate a simple extension to
\Thref{WernersepTh} to characterize mixtures of tripartite product
states\\

\noindent\textbf{Theorem \ref{WernersepTh}'} \textit{(Three-party
Separability)\\ A Gaussian three-party state with CM $\gamma$ can be
written as a mixture of tripartite product states iff there exist
one-mode correlation matrices $\gamma_A, \gamma_B,
\gamma_C$ such that
\begin{equation}\label{fullsepdef}
\gamma-\gamma_A\oplus\gamma_B\oplus\gamma_C\geq 0.
\end{equation}
Such a state will be called \emph{fully separable}.
}

\noindent\Proof\ The proof is in complete analogy with that of
\Thref{Wernersep} in \cite{GBE} and is therefore omitted here.

A state for which there are a one-mode CM $\gamma_A$ and a two-mode CM
$\gamma_{BC}$ such that $\gamma-\gamma_A\oplus\gamma_{BC}\geq 0$ is
called $A-BC$ \emph{biseparable} (and similarly for the two other
bipartite groupings).
In total, we have the following five different entanglement classes:

\begin{description}
\item[Class 1] \emph{Fully inseparable states} are those which
are not separable for any grouping of the parties.
\item[Class 2] \emph{1-mode biseparable states} are those which are
separable if two of the parties are grouped together, but
inseparable with respect to the other groupings.
\item[Class 3] \emph{2-mode biseparable states} are separable with
respect to two of the three bipartite splits but inseparable with
respect to the third.
\item[Class 4] \emph{3-mode biseparable states} separable with respect
to all three bipartite splits but cannot be written as a mixture of
tripartite product states.
\item[Class 5] The \emph{fully separable} states can be written as a
mixture of tripartite product states.
\end{description}

Examples for Class 1 (the GHZ-like states of \cite{CVghz}), Class 2
(two-mode squeezed vacuum in the first two and the vacuum in the third
mode), and Class 5 (vacuum state in all three modes) are readily
given; we will provide examples for Classes 3 and 4 in Subsection
\ref{pptesSec} below.

How can we determine to which Class a given state with CM $\gamma$
belongs?  States belonging to Classes 1, 2, or 3 can be readily
identified using the NPT-criterion (\Thref{nptcrit}). Denoting the
partially transposed CM by $\tilde\gamma_x = \Lambda_x\gamma\Lambda_x,
x=A,B,C$, we have the following equivalences:
\begin{lemma} (Classification)\label{classific}
\begin{eqnarray}
\tilde\gamma_A\not\geq iJ,
\tilde\gamma_B\not\geq iJ,
\tilde\gamma_C\not\geq iJ
&\Leftrightarrow&\mbox{Class 1}\\
(*) \tilde\gamma_A\not\geq iJ,
\tilde\gamma_B\not\geq iJ,
\tilde\gamma_C\geq iJ &\Leftrightarrow&\mbox{Class 2}\\
(*) \tilde\gamma_A\not\geq iJ,
\tilde\gamma_B\geq iJ,
\tilde\gamma_C\geq iJ &\Leftrightarrow&\mbox{Class 3}\\
\tilde\gamma_A\geq iJ,
\tilde\gamma_B\geq iJ,
\tilde\gamma_C\geq iJ &\Leftrightarrow&\mbox{Class 4 or 5},\label{openq}
\end{eqnarray}
where the $(*)$ reminds us to consider all permutations of the indices
A, B, and C.
\end{lemma}
The proof follows directly from the definitions of the different
classes and \Thref{nptcrit}.

What is still missing is an easy way to distinguish between Class 4
and Class 5. Thus to complete the classification we now provide a
criterion to determine whether a CM $\gamma$ satisfying
\Ineqsref{openq} is fully separable or 3-mode biseparable, that is we
have to decide whether there exist one-mode CMs
$\gamma_A,\gamma_B,\gamma_C$ such that (\ref{fullsepdef}) holds, in
which case $\gamma$ is fully separable.  In the next subsection we
will describe a small set consisting of no more than nine CMs among
which $\gamma_A$ is necessarily found if the state is separable.

\subsection{Criterion for Full Separability}

This subsection contains the main result of the paper: a separability
criterion for PPT $1\times1\times1$ Gaussian states, i.e. states whose
CM fulfills \Ineqsref{openq}. We start from \Thref{WernersepTh}' and
obtain in several steps a simple, directly computable necessary and
sufficient condition. The reader mainly interested in this result may
go directly to \Thref{fullsepTh}, from where she will be guided to the
necessary definitions and Lemmas.

Since the separability condition in \Thref{WernersepTh}' is formulated
in terms of the positivity of certain matrices the following lemma
will be very useful throughout the paper. We
consider a self-adjoint $(n+m)\times(n+m)$ matrix $M$ that we write in
block form as
\begin{equation}\label{block}
M = \left( \begin{array}{cc} A&C\\C^\dagger&B\end{array} \right),
\end{equation}
where $A, B, C$ are $n\times n, m\times m$, and $n\times m$ matrices,
respectively.
\begin{lemma} (Positivity of self-adjoint matrices)\\ \label{positivity}
A self-adjoint matrix $M$ as in (\ref{block}) with $A\geq0,
B\geq0$ is positive if and only if
for all $\epsilon>0$
\begin{equation}\label{posconda}
A-C\frac{1}{B+\epsilon\id}C^\dagger\geq0,
\end{equation}
or, equivalently, if and only if
\begin{subequations}
\begin{equation}\label{kernelcond}
\ker B\subseteq\ker C
\end{equation}
and
\begin{equation}\label{poscondb}
A-C\frac{1}{B}C^\dagger\geq0,
\end{equation}
\end{subequations}
where $B^{-1}$ is
understood in the sense of a pseudoinverse (inversion on the range).
\end{lemma}
\Proof\ The only difficulty in the proof arises if $\ker B\not={0}$.
Therefore we consider the matrices $M_\epsilon$, where $B$ in (\ref{block})
is replaced by $B_\epsilon=B+\epsilon\id$ ($\epsilon>0$), which avoid this
problem and which are positive $\forall\epsilon>0$ iff $M\geq0$. In a second
simplifying step we note that $M_\epsilon\geq0$ $\forall\epsilon>0$ iff
$M'_\epsilon=(\id\oplus B_\epsilon^{-1/2})M(\id\oplus
B_\epsilon^{-1/2})\geq0$.

Now direct calculation shows the claim: we can write a general
$f\oplus g$ as
$f\oplus\left[(B_\epsilon^{-1/2}C^\dagger)h+h_\bot\right]$, where
$h_\bot$ is orthogonal to the range of
$(B_\epsilon^{-1/2}C^\dagger)$. Then $(f\oplus g)^\dagger
M'_\epsilon(f\oplus g) =
f^\dagger(A-CB_\epsilon^{-1}C^\dagger)f+(f+h)^\dagger
CB_\epsilon^{-1}C^\dagger(f+h)+h_\bot^\dagger h_\bot$, which is
clearly positive, if (\ref{posconda}) holds. With the choice
$h_\bot=0$ and $h=-f$ it is seen that (\ref{posconda}) is also
necessary.

That the second condition is equivalent is seen as follows:
If \Ineqref{posconda} holds $\forall\epsilon>0$ there cannot be
vector $\xi\in\ker B$ and $\xi\not\in\ker C$ since for such a $\xi$ we
have $\xi^T\left(A-C\frac{1}{B+\epsilon\id}C^\dagger \right)\xi<0$ for
sufficiently small $\epsilon>0$, and if (\ref{kernelcond}) holds then
(\ref{posconda}) converges to (\ref{poscondb}). Conversely, if
(\ref{kernelcond}) holds, then $CB^{-1}C^\dagger$ is well-defined and
\Ineqref{poscondb} implies it $\forall\epsilon>0$.\hfill\qed

As mentioned above, in this section  we exclusively consider
three-mode CMs $\gamma$ that satisfy \Ineqsref{openq}. We write
$\gamma$ in the form of \Eqref{block} as
\begin{equation}\label{gamma3}
\gamma=\left( \begin{array}{cc}A&C\\ C^T&B
\end{array} \right),
\end{equation}
where $A$ is a $2\times 2$ matrix, whereas $B$ is a $4\times 4$ matrix. We
observe that \Ineqsref{openq} impose some  conditions on $\gamma$ that will be
useful later on:
\begin{observation}\label{obs1}
Let $\gamma$ satisfy \Ineqsref{openq}, then
\begin{equation}
\gamma\geq\left( \begin{array}{ccc} \sigma_A iJ&0&0\\
0&\sigma_B iJ&0\\
0&0&\sigma_C iJ\\
\end{array} \right),
\end{equation}
where $\sigma_x\in\left\{ 0,\pm1 \right\},\forall x=A,B,C$.
\end{observation}
\Proof\ \Ineqsref{openq} say that $\gamma\pm iJ\geq0$ and $\gamma\pm i\tilde
J_x\geq0$ $\forall x$. By adding these positive matrices all combinations of
$\sigma_x$ can be obtained.\hfill\qed

From this it follows:
\begin{observation}\label{obs2}
For a PPT CM $\gamma$ as in \Eqref{gamma3}
\begin{equation}\label{kernelcond2}
\ker(B+iJ), \ker(B+i\tilde J)\subseteq\ker C,
\end{equation}
where $\tilde J=J\oplus(-J)$ is the partially transposed $J$ for two
modes.
\end{observation}
\Proof\ Cond.\ (\ref{kernelcond2}) on the kernels is an immediate
consequence of \Leref{positivity} applied to the matrices
$\gamma-0\oplus iJ\oplus (\pm iJ)$, which are positive by
Obs. \ref{obs1}. \hfill\qed

Then the matrices
\begin{subequations}\label{Ndef}
\begin{eqnarray}
\tilde{N}&\equiv& A-C\frac{1}{B-i\tilde
J}C^T,\\
N&\equiv& A-C\frac{1}{B-i J}C^T
\end{eqnarray}
\end{subequations}
are well-defined and

\begin{observation}\label{obs2b}
It holds that both
\begin{equation}\label{trcond}
\tr N, \tr\tilde N > 0,
\end{equation}
\end{observation}
\Proof\ Cond.\ (\ref{trcond}) is true since, again by
\Leref{positivity} and Obs.\ \ref{obs1}, both $N$ and $\tilde N$ are
positive and $N\pm iJ, \tilde N\pm iJ\geq 0$.  This implies that $N,
\tilde N$ cannot be zero, which is the only positive matrix with
vanishing trace. Therefore $\tr N, \tr\tilde N$ are strictly
positive. \hfill\qed

The remainder of this section leads in several steps to the
separability criterion.
First, we simplify the condition (\ref{fullsepdef}) by reducing it to
a condition which involves only one one-mode CM $\gamma_A$.

\begin{lemma}\label{fslemma}
A PPT 3-mode CM $\gamma$ is fully  separable if and only if there exists a
one-mode CM $\gamma_A$ such that
\begin{subequations}\label{fullsepb}
\begin{eqnarray}
\label{fullsepb1}
\tilde{N}&\geq&\gamma_A,\\
\label{fullsepb2}
N&\geq&\gamma_A,
\end{eqnarray}
\end{subequations}
where $N, \tilde N$ were defined in \Eqsref{Ndef}. Without loss of
generality we require $\gamma_A$ to be a pure state CM, i.e.\
$\det\gamma_A=1$.
\end{lemma}
\Proof\ By \Thref{WernersepTh}' full separability of $\gamma$ is
equivalent to the existence of one-mode CMs
$\gamma_A,\gamma_B,\gamma_C\geq iJ$ such that
$\gamma-\gamma_A\oplus\gamma_B\oplus\gamma_C\geq0$.  Let $\gamma_x$
stand for $\gamma_{A,B,C}$.

By \Leref{positivity} this is equivalent to $\exists\gamma_x$ such that
$X_\epsilon\equiv B - C^T\frac{1}{A_\epsilon-\gamma_A}C \geq
\gamma_B\oplus\gamma_C$, $\forall\epsilon>0$, where $A_\epsilon\equiv
A+\epsilon\id$. But iff there exist such $\gamma_x$ then (\Leref{positivity})
the inequality also holds for $\epsilon=0$ and the kernels fulfill
(\ref{kernelcond}). This is true iff the matrix $X\equiv X'_0$ is a CM
belonging to a separable state, i.e.\ (\Thref{nptcrit}) iff $X'\geq i\tilde
J, iJ$. Using $B\geq i\tilde J, iJ$ [which holds since $\gamma$ fulfills
\Ineqsref{openq}] we obtain that $\gamma$ is separable iff there exists
$\gamma_A\geq iJ$ such that
\begin{equation}\label{step1}
\left( \begin{array}{cc} A-\gamma_A&C\\ C^T&B'_k
\end{array} \right)\geq 0,\,\, k=1,2,
\end{equation}
where $B'_1=B-iJ$ and $B'_2=B-i\tilde J$.  Since Condition \
(\ref{kernelcond}) holds, this is (\Leref{positivity}) equivalent to
\Ineqsref{fullsepb}. That we can always choose $\det\gamma_A=1$ follows
directly from \Eqref{iv} and the remark after \Leref{physical}.\hfill\qed

While we can always find a $\gamma_A$ fulfilling \Ineqref{fullsepb2}, since
$\gamma$ belongs to a PPT state (and there exists a two-mode CM
$\gamma_{BC}\geq iJ$ such that $\gamma_A\oplus\gamma_{BC}$ is smaller than
$\gamma$), it may well happen that \Ineqref{fullsepb1} cannot be satisfied at
all, or that it is impossible to have both \Ineqsref{fullsepb} fulfilled for
one $\gamma_A$ simultaneously. Note that due to \Ineqsref{openq}, $N$ and
$\tilde N$ as above are always positive. From \Ineqsref{fullsepb} we observe
that
\begin{observation}\label{obsnecsep}
for the CM $\gamma$ of a separable state it is necessary  to have
\begin{subequations}\label{neccond}
\begin{eqnarray}
\label{necconda}
\tr N, \tr\tilde N&\geq& 2,\\
\label{neccondb}
\det N, \det\tilde N &>& 0,
\end{eqnarray}
\end{subequations}
where $\gamma$ as in \Eqref{gamma3} and  $N, \tilde N$ as in
\Eqsref{Ndef}.  
\end{observation}
\Proof\ A self-adjoint $2\times 2$ matrix is positive iff its trace and
determinant are positive. Since the trace of the RHS of both
\Ineqsref{fullsepb} is $\geq2$ [remark after \Leref{physical}] the same is
necessary for the LHS. Also, since $\det\gamma_A=1$, which implies that $\gamma_A$ has
full rank, any matrix $\geq\gamma_A$ must also have full rank \cite{2xNpaper}
and thus a strictly positive determinant.\hfill\qed

For a self-adjoint positive $2\times2$ matrix
\begin{equation}\label{N}
R = \left(
\begin{array}{cc}a&b\\b^*&c\end{array} \right),
\end{equation}
we show
\begin{lemma}\label{smallerlem}
There exists a CM $\gamma_A\leq R$ if and only if there exist
$(y,z)\in\RR^2$ such that
\begin{subequations}\label{smaller}
\begin{eqnarray}
\label{smallera}
\tr R &\geq& 2\sqrt{1+y^2+z^2},\\
\label{smallerb}
\det R +1 + L^T{y \choose z} &\geq& \tr R \sqrt{1+y^2+z^2},
\end{eqnarray}
\end{subequations}
where
\begin{equation}\label{Ldef}
L = (a-c, 2\Re b).
\end{equation}
\end{lemma}
\Proof\ As noted in \Leref{fslemma} we need only look for $\gamma_A$
with $\det\gamma_A=1$. We parameterize
\begin{equation}\label{onemode}
\gamma_A = \left( \begin{array}{cc} x+y&z\\
z&x-y\end{array} \right),
\end{equation}
with real parameters $x,y,z$ and $x^2=1+y^2+z^2$ for purity.
This is a CM iff $\gamma_A-iJ\geq0$ (\Leref{physical}), that
is iff $\tr \gamma_A=2x\geq0$ (where we use that positivity of the a
$2\times2$ matrix is
equivalent to the positivity of its trace and determinant and
$\det(\gamma_A-iJ)=0$ by construction). By the same
argument, $R-\gamma_A\geq0$ leads to the two conditions
(\ref{smaller}). \hfill\qed

The inequalities (\ref{smaller}) have a simple geometrical
interpretation that will be useful for the proof of the promised
criterion: \Ineqref{smallera} restricts $(y,z)$ to a
circular disk $\cC'$ of radius $\sqrt{(\tr
R)^2/4-1}$ around the origin, while
\Ineqref{smallerb} describes a (potentially degenerate) ellipse
$\cE$ (see \Figref{circles}), whose elements are calculated below,
and the existence of a joint solution to \Ineqsref{smaller} is therefore
equivalent to a nonempty intersection of $\cC'$ and $\cE$.

Applying this now to the matrices (\ref{Ndef}) we find that in order to
simultaneously satisfy both conditions in \Leref{fslemma}, the intersection
between the two ellipses $\cE, \tilde\cE$ and the smaller of the two
concentric circles $\cC',\tilde\cC'$ (which we denote in the following by
$\cC$) must be nonempty. This condition leads to three inequalities in the
coefficients of the matrices $\tilde N, N$ which can be satisfied
simultaneously if and only if the PPT trimode state is separable. Thus we can
reformulate the condition for separability (\Leref{fslemma}) as follows
\begin{lemma}\label{reformsepcrit} (Reformulated Separability
Condition)\\
A three-mode state with CM $\gamma$ satisfying \Ineqsref{openq} is
fully separable if and only if
there exists a point $(y,z)\in\RR^2$
fulfilling the following inequalities:
\begin{subequations}\label{sepcond}
\begin{eqnarray}
\label{scc}
\mathrm{min}\{\tr N, \tr \tilde{N}\} &\geq&
2\sqrt{1+y^2+z^2},\\
\label{sce}
 \det N +1 + L^T{y \choose z}
&\geq& \tr N \sqrt{1+y^2+z^2},\\
\label{scet}
 \det \tilde{N} +1 +
\tilde{L}^T{y \choose z} &\geq& \tr \tilde{N}
\sqrt{1+y^2+z^2}.
\end{eqnarray}
\end{subequations}
\end{lemma}
\Proof\ According to \Leref{fslemma} $\gamma$ belongs to a separable
state iff we can find $\gamma_A$ smaller than $\tilde N$ and smaller
than $N$. According to \Leref{smallerlem} we can find such a
$\gamma_A$  iff we can find $(y,z)$ such that \Ineqsref{smaller} are
satisfied for both $N$ and $\tilde N$.\hfill\qed

In the following paragraphs we have a closer look at the sets $\cE,
\tilde\cE$, and $\cC$. The goal of this discussion is to identify a
few special points -- directly computable from $\gamma$ -- among which
a solution to \Ineqsref{sepcond} will be found iff the state under
consideration is separable. This will then lead to the final practical
form of the separability criterion which is stated at the end of this
section.

By squaring \Ineqref{sce} we obtain
\begin{equation}\label{ellipse}
\left[{y\choose z}-\mu L \right]^T K \left[{y\choose z}-\mu L \right]
\leq m,
\end{equation}
where $\mu = (\det N +1)/k_1$, $m =
\frac{k_2}{k_1}\left[ (\det N +1)^2 -k_1\right]$, and the
matrix $K$ is \cite{fnLnot0}
\[
K = k_1P_L + k_2P_{L^\bot},
\]
with the orthogonal
projectors $P_L, P_{L^\bot}$ on $L,L^\bot$ and
\begin{eqnarray*}
k_1 & = & 4\left[\det N +(\Im b)^2\right],\\
k_2 & = & (\tr N)^2.
\end{eqnarray*}
Due to \Ineqsref{neccond} $k_1$ and $k_2$ are strictly positive, $\mu, m$ are
well-defined and $K$ is a positive matrix of rank $2$. Let us now distinguish
the cases $m<0$ and $m\geq0$. For $m<0$ \Ineqref{ellipse} can never be
fulfilled since $K$ is a positive matrix. In the case $m\geq0$,
\Ineqref{ellipse} describes an ellipse $\cE$ which is centered at $m_e = \mu
L$ with major axis $L$ and minor axis $L^\bot$ of lengths
$\sqrt{m/k_1}\geq\sqrt{m/k_2}$, respectively. From \Ineqref{scet} we obtain
the same equations for the tilded quantities derived from $\tilde N$.

The final argument for the derivation of the separability
criterion is as follows. By \Leref{reformsepcrit} the state is
separable if and only if the three sets described by
\Ineqsref{sepcond} have a common intersection, i.e.\ iff $I \equiv
\cE\cap\tilde\cE\cap\cC\not=\emptyset$.  The border of $I$ is
contained in the union of the borders of the ellipses and circle:
$\del I\subseteq\del\cE\cup\del\tilde\cE\cup\del\cC$.  Now we can
distinguish two cases, both of which allow to calculate a definite
solution to the \Ineqsref{sepcond} if the state is separable: Either
$\del I$ has nonempty intersections with the borders of two of the
sets $\cE, \tilde\cE, \cC$ or $\del I$ coincides with the border of
one of the three. In the latter case this whole set is contained in
$I$. In the former case, at least one of the points at which the
borders intersect must be in $I$ and thus a solution. If no solution
is found this way the state is inseparable.  This argument is made
more precise in the final theorem. Formulas for the nine candidate
solutions -- the centers $m_c, m_e, m_{\tilde e}$ and the
intersections points $i^\pm_{e\tilde e}, i^\pm_{ce}, i^\pm_{c\tilde
e}$ -- are given in
Appendix \ref{app}.

\begin{theorem} (Criterion for full separability)\\ \label{fullsepTh}
A three-mode state corresponding to the CM
$\gamma$ satisfying \Ineqref{openq} is fully separable if and only if
\Ineqref{neccondb} holds and there exists a point $\xi_{sol}$
\begin{equation}\label{points}
\xi_{sol}\in \{ m_c,m_e,
m_{\tilde{e}},
i_{e\tilde{e}}^{\pm},i_{ce}^{\pm},i_{c\tilde e}^{\pm}\}
\end{equation} fulfilling the \Ineqsref{sepcond}.
\end{theorem}

\Proof\  We already saw (Obs. \ref{obsnecsep}) that $\det N,
\det\tilde N>0$ are necessary for separability. If this holds, the
quantities used in (\ref{sepcond}, \ref{points}) and in their
derivation are all well-defined.

According to \Leref{reformsepcrit} $\gamma$
is fully separable iff there exists a point
$(y,z)^T$ such that the \Ineqsref{sepcond} are fulfilled.
Therefore, if one of the points (\ref{points}) satisfies
\Ineqsref{sepcond} then it determines a $\gamma_A$ fulfilling
\Ineqsref{fullsepb} thus proving that the state is separable.
To complete the proof, we show that if the state is separable, then we
find a solution to \Ineqsref{sepcond}
among the points (\ref{points}).

As pointed out before, the condition that \Ineqsref{sepcond} can
simultaneously be satisfied has the geometrical
interpretation that the circle $\cC$ and the two ellipses
$\cE,\tilde\cE$ have a nonempty intersection,
i.e.\ $I \equiv \cE\cap\tilde\cE\cap\cC\not=\emptyset$.

Thus it remains to prove that if $I$ is nonempty then one of the nine
points in (\ref{points}) lies in $I$. But if $I\not=\emptyset$
there are only the following two possibilities: since all the sets
considered are convex and closed, either the border of $I$
coincides with that of one of the sets $\cC, \cE, \tilde\cE$ (which
means that
one of these sets, call it $\cS$, is contained in \emph{both} others)
or at least two of the borders $\del\cC, \del\cE, \del\tilde\cE$
contribute to $\del I$, in which case the points at which these two
intersect belong to $\del I$ and thus to $I$.

In the former case, the center of $\cS$ is a solution and given by one
of the \Eqsref{fullsep1}; in the latter, one can find a solution among
the intersections of the borders of the sets $\cE,\tilde\cE,\cC$. That
these are given by the $i^\pm_x$ is shown in Appendix \ref{app}.\hfill\qed

If a CM $\gamma$ belongs to a separable state according to the above
theorem then the point $\xi_{sol}$ provides us with a pure one-mode CM
$\gamma_A$ such that $N, \tilde N\geq\gamma_A$. By construction
$\gamma'=B-C (A-\gamma_A)^{-1}C^T$ is a separable $2\times2$ CM and by
repeating a similar procedure as above with $\gamma'$ we can calculate
a pure product-state decomposition of the original state with CM
$\gamma$.

\section{Examples of Bound Entangled States}\label{pptesSec}
In this section we construct states belonging to Classes 3 and
4. Our construction makes use of ideas that were first applied
 in finite dimensional quantum systems to find PPT entangled states
(PPTES) \cite{pptes} and then generalized in \cite{edge} to construct
so-called \emph{edge states}, i.e.\ states on the border of the convex set of
states with positive partial transpose.  Similarly, one can define ``edge
CMs'' as those that lie on the border of the convex set of PPT CMs (they are
called ``minimal PPT CMs'' in \cite{GBE}).

This section is divided into three subsections. In the first one
we define ``edge CMs'' and characterize them. In the second and
third subsections we present two different families of CMs which
contain edge CMs. We also show that within those families we
have CMs belonging to all classes.

\subsection{Edge CMs}\label{edgeCMs}
In the following we will consider CMs $\gamma$ corresponding to PPT states,
i.e. fulfilling
\begin{equation}
\label{cond1}
\gamma - i \tilde J_x\ge 0, \quad \hbox{ for all } x=0,A,B,C,
\end{equation}
where $\tilde J_0\equiv J$.
\begin{definition} (Edge Correlation Matrices)\\
\label{edgeCMdef}
A CM $\gamma$ is an edge CM if it
corresponds to a non--separable state, fulfills (\ref{cond1}),
and $\gamma'\equiv \gamma - P$ does not fulfill (\ref{cond1})
for all real operators $P$ with $0\ne P\ge 0$.
\end{definition}

Note that a state with an edge CM automatically belongs to class 4
(i.e.\ edge CMs correspond to 3-mode biseparable states). In order to
fully
characterize them, we will need the following definition. Let us
consider the complex vector space $V\subseteq \CC^6$ of dimension $d$
spanned by
the vectors belonging to the kernels of all $\gamma-i\tilde J_x$
($x=0,A,B,C$). We will define $K(\gamma)$ as a real vector space
which is spanned by the real parts and imaginary parts of all
the vectors belonging to $V$. More specifically, let us denote
by $B=\{f_R^k + i f_I^k\}_{k=1}^d$ a basis of $V$, such that
$f_R^k$ and $f_I^k$ are real. We define
\begin{equation}
\label{K}
K(\gamma)=\left\{\sum_k \lambda_k f_R^k + \mu_k f_I^k,
\lambda_k,\mu_k\in \RR\right\}\subseteq \RR^6;
\end{equation}
that is, the real span of the vectors $f_R^k$ and $f_I^k$. Note
that this definition does not depend on the chosen basis $B$ [As
it is pointed out in Appendix \ref{app2}, $K(\gamma)$ coincides with
the real vector space spanned by all the vectors in the kernels
of $\gamma+\tilde J_x\gamma^{-1}\tilde J_x$]. We then have the
following

\begin{theorem} (Characterization of $1\times1\times1$ edge CMs)\\
A CM $\gamma$ fulfilling (\ref{cond1}) is an edge CM if and only if
there exist no CMs $\gamma_A,\gamma_B,\gamma_C$ such that
$\gamma=\gamma_A \oplus\gamma_B\oplus\gamma_C$ and $K=\RR^6$.
\end{theorem}

\Proof\ We will use the fact \cite{2xNpaper} that, given two
positive matrices $A,B\ne 0$, there exists some $\epsilon>0$
such that $A-\epsilon B\ge 0$ iff ran($B$)$\subseteq$ ran($A$).
According to the Def. \ref{edgeCMdef} we cannot
subtract any real positive matrix from $\gamma$ without
violating the conditions (\ref{cond1}). This is equivalent to
imposing that there is no real vector in the intersection of the
ranges of the matrices $\gamma-i\tilde J_x$. This is again
equivalent to saying that there is no real vector orthogonal to
all the ker($\gamma-i\tilde J_x)$, which in turn is equivalent
to $K=\RR^6$, since that vector should be orthogonal to all the
real and imaginary parts of the vectors spanned by those
kernels. Now, if $\gamma$ corresponds to an entangled state it
is clear that $\gamma\not=\gamma_A
\oplus\gamma_B\oplus\gamma_C$. Conversely, if $\gamma\not=\gamma_A
\oplus\gamma_B\oplus\gamma_C$ was separable, then there must exist some
real positive $P$ such that $\gamma-P=\gamma_A
\oplus\gamma_B\oplus\gamma_C$ is separable, and therefore
fulfills (\ref{cond1}), which is not possible.
\hfill\qed

Note that this Theorem generalizes easily to the cases of more than
three parties and more than one mode at each site. 

In the construction of the following two examples of tripartite
bound entangled states we are going to use this theorem. The
idea is to take a CM $\gamma_0$ of a pure entangled state
[which, of course, does not fulfill (\ref{cond1})] and add real
positive matrices until the conditions (\ref{cond1}) as well as
$K=\RR^6$ are fulfilled. If the resulting CM is not of the form
$\gamma_A \oplus\gamma_B\oplus\gamma_C$ then Theorem 4 implies
that it is an edge CM. In fact, we can add more real positive
matrices keeping the state entangled [and fulfilling
(\ref{cond1})]. In order to see how much we can add, we can use
the criterion derived in the previous section.

This method of constructing CMs belonging to Class 4 also
indicates how the corresponding states may be prepared
experimentally. Adding a positive matrix $P$ to the CM
$\gamma_0$ corresponds to the following preparation process:
start with an ensemble of states with CM $\gamma_0$, displace
them randomly by $d$ according to the Gaussian probability
distribution with covariance matrix given by the inverse of $P$. This
is a  \emph{local} operation (that potentially needs to be
supplemented by classical communication) on each individual
mode. The state produced by this randomization has CM $\gamma+P$
\cite{GBE}.

\subsection{Example 1}
\label{extmss}

In the first example we start out with an entangled state
between the two parties Alice and Bob and the vacuum state in
Charlie and add two projectors to the corresponding CM. More
specifically, we consider the CMs of the form
$\gamma_{a_1,a_2}=\gamma+a_1P_1+a_2P_2$, where
\begin{equation}
\gamma=\gamma_{AB}\oplus \id_C,
\end{equation}
and
\begin{equation}
\gamma_{AB}= \left(\begin{array}{cccc} a&0&c&0\\ 0&a&0&-c\\
c&0&a&0\\0&-c&0&a\end{array}\right),
\end{equation}
with $a=\sqrt{1+c^2}$ and $c$ can take any value different from
zero. Here, $P_1=\tilde{p}_1
\tilde{p}_1^T$ and $P_2=\tilde{p}_1
\tilde{p}_1^T$, where $\tilde{p}_1=(0,1,0,1,1,2)^T$ and
$\tilde{p}_2=(1,0,-1,0,0,1)^T$.

In order to explain why the CM $\gamma_{a_1,a_2}$ achieves our
purposes, let us first consider the two--mode case in which the
correlation matrix is $\gamma_{AB}$. We denote now by
$p=p_1+ip_2$ [where $p_1=(0,1,0,1)^T$ and $p_2=(1,0,-1,0)^T$]
the eigenvector corresponding to the negative eigenvalue of
$\gamma_{AB}-i\tilde{J}_A$ \cite{fndimofJ}. Since
$(-i\tilde{J}_A)^\ast=-i\tilde{J}_B$ we have that the
eigenvector corresponding to the negative eigenvalue of
$\gamma_{AB}-i\tilde{J}_B$ is $p^\ast=p_1-ip_2$. By adding a
sufficiently large multiple of the projectors onto those
vectors, we obtain a CM whose partial transposes are positive.
Note that in this case (just two modes) this would already make
the state separable.

In the case of three modes with a correlation matrix $\gamma$
the same argumentation applies, namely that by adding some
projectors we can make the partial transposes with respect to A
and B positive. However, we have to involve $C$ and thereby
smear out the initial entanglement between $A$ and $B$ among all
three parties. This is exactly what is achieved by adding the
projectors $P_1$ and $P_2$. If we choose now, for instance,
$c=0.3$, $a_1=1$, and $a_2\approx 0.5531095$, then one can
show that the set $K(\gamma_{a_1,a_2})$ defined as in \Eqref{K}
spans $\RR^{6}$. As mentioned at the end of the previous
subsection, since the resulting CM is not of the form
$\gamma_A\oplus\gamma_B \oplus \gamma_C$ it corresponds to an
edge CM.

\begin{figure}[tbp]
\epsfxsize=8.5cm
\epsfysize=6.6cm
\begin{center}
\epsffile{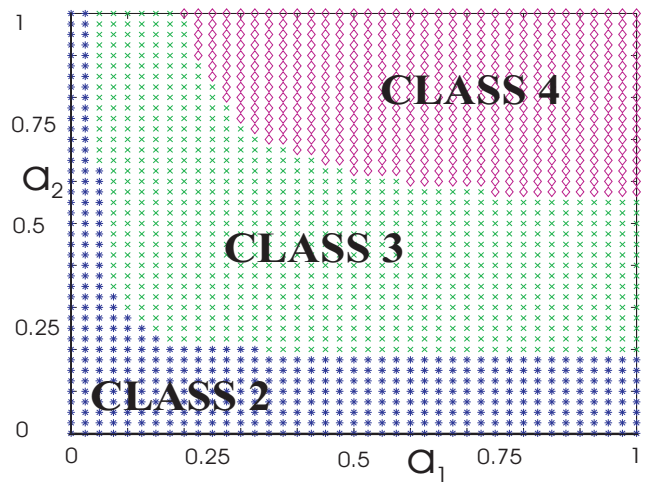}
\caption{The Entanglement Classes of
$\gamma_{a_1,a_2}$\label{extmssfig}} 
\end{center}
\end{figure}

In \Figref{extmssfig} we illustrate to which class
$\gamma_{a_1,a_2}$ belongs as a function of the parameters
$a_{1,2}$. In order to determine this, we have used the
criterion derived in the previous section. It is worth noting
that $\gamma_{a_1,a_2}$ never becomes separable. This follows
from \Thref{fullsepTh} and the fact that both $m=\tilde{m}=0$
for all values of $a_{1,2}$, as can be easily verified. This
implies that the two ellipses [c.f.\ \Ineqref{ellipse}] are
just two points [which coincide with the centers given in
\Eqref{fullsep1}]. Thus, the only possibility that the circle
and the two ellipses intersect is that the centers of the
ellipses are the same and lie inside the circle. It is easy to
show that for all values of $a_1$ and $a_2$ the centers of the
two ellipses are never the same. Thus the state corresponding to
the CM $\gamma_{a_1,a_2}$ is never separable and is a PPTES for
all values of $a_1$, $a_2$ for which the partial transposes are
positive.

\subsection{Example 2}
\label{exghz}

Here we present a family of states which belong either to Class
1, 4, or 5. The states of this family are obtained from a pure
GHZ-like state \cite{CVghz} by adding a multiple of the identity,
i.e.,
\begin{equation}
\gamma_\alpha=\gamma+\alpha\id,
\end{equation}
where
\begin{equation}
\gamma =\left(
\begin{array}{cccccc}
a&0&c&0&c&0\\ 0&b&0&-c&0&-c\\ c&0&a&0&c&0\\ 0&-c&0&b&0&-c\\
c&0&c&0&a&0\\0&-c&0&-c&0&b
\end{array}\right),
\end{equation}
with $a>1$ and
\begin{eqnarray}
b=\frac{1}{4}(5a-\sqrt{9a^2-8}),\\
c=\frac{1}{4}(a-\sqrt{9a^2-8}).
\end{eqnarray}

For the following discussion, we pick $a=1.2$.
It is clear that for $\alpha=0$ the state is fully inseparable,
i.e., it belongs to Class $1$, whereas for $\alpha\geq 1$ the
state will be fully separable (Class $5$). We will show now that
for $\alpha_0\leq
\alpha \leq \alpha_1$, where $\alpha_0\approx 0.29756$ and
$\alpha_1\approx 0.31355$ the state is biseparable and belongs
therefore to Class $4$.

The CM $\gamma_\alpha$ is symmetric with respect to permutations
between the parties, and therefore the negative eigenvalues of
the matrices $\gamma-i\tilde J_x, x=A, B, C$ are the same. We
denote its absolute value by $\alpha_0\approx 0.29756$. It is
easy to determine the real and imaginary part of the
corresponding eigenvectors. One finds that all those vectors are
linearly independent. If we add now $\alpha_0 \id$ to $\gamma$
then all those vectors belong to $K(\gamma_{\alpha_0})$ which
immediately implies that $K(\gamma_{\alpha_0})=\RR^6$. Since
$\gamma_{\alpha_0}\not=\gamma_A \oplus\gamma_B\oplus\gamma_C$ we
have that it is an edge CM.

\begin{figure}[tbp]
\epsfxsize=11cm
\epsfysize=5cm
\begin{center}
\epsffile{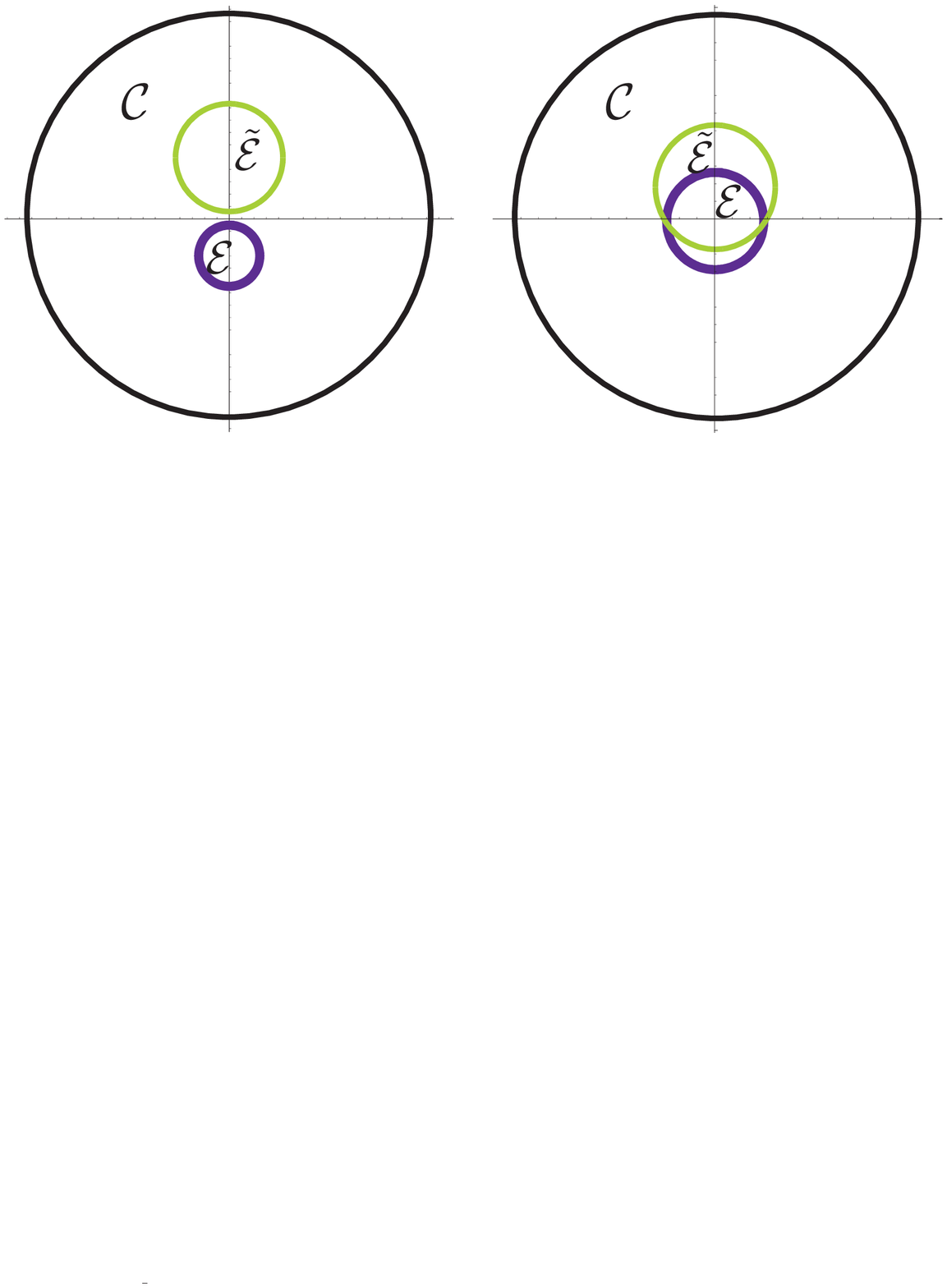}
\caption{(a) The circle and the two ellipses do not have a joint
intersection, therefore the state corresponding to $\gamma_\alpha$
is a PPTES, (b) the circle and the two ellipses have a joint
intersection, therefore the state corresponding to $\gamma_\alpha$ is
separable. \label{circles}} 
\end{center}
\end{figure}

Let us now use Theorem $3$ in order to determine $\alpha_1$.
First of all, we show, independently of the discussion above
that $\gamma_{\alpha_0}$ belongs to Class 4. In particular, we
find that $m=\tilde m = 0$ [cf. \Eqref{ellipse}], which implies
that there exists a solution to the \Ineqsref{sepcond} only if
the centers of the two ellipses are the same and lie within the
circle. Here one can also show that the two centers are not the
same and so the state corresponding to the CM
$\gamma_{\alpha_0}$ is a PPTES. Let us determine the values of
$\alpha$ for which it is still the case that there exists no
intersection of the two ellipses and the circle given by the
\Ineqsref{sepcond}. It is easy to show that if $\alpha>\alpha_0$
then $\tr N \leq \tr \tilde N$, which implies that the circle
that has to be considered has radius $r_c=\sqrt{(\tr N)^2/4-1}$.
One can also easily verify that the two ellipses never intersect
the border of the circle, which simplifies the problem. The
ellipses must always lie inside the circle (since if they where
outside it would never be possible to obtain a separable state
even for $\alpha>1$). Thus, the problem reduces to check at
which point the ellipses intersect each other. This occurs when
$\alpha=\alpha_1\approx 0.31355$. Thus the CM $\gamma_\alpha$,
where $\alpha_0 \leq \alpha < \alpha_1$ corresponds to a PPTES,
whereas for $\alpha\geq \alpha_1$, the corresponding state is
fully separable. In \Figref{circles} we have
plotted the circle and the two ellipses, which are almost
circles in this case, for (a) $\alpha <\alpha_1$ and (b) $\alpha
>\alpha_1$.

\section{Conclusions}
We have discussed nonlocal properties of Gaussian states of three
tripartite modes. We have distinguished five classes with different
separability properties and given a simple necessary and sufficient
criterion that allows to determine which of these classes a given
Gaussian state belongs to. The first three classes contain only NPT
states and positivity of a state under the three partial
transpositions suffices to determine to which of those it belongs. The
separability criterion, which allows to distinguish PPT entangled
states from separable states is the main result of this paper. For the
case of three qubits such a criterion is still missing. Lastly, we
have constructed examples for all the classes and in particular for
tripartite entangled state with positive partial transpose. Using the
separability criterion for multi-mode bipartite Gaussian states
\cite{GSepCrit} the results presented above can be extended to cover
the case of $n$ modes at location C. Nothing changes in the argumentation
to distinguish 3-party biseparable from fully separable states [the
additional modes are taken care of automatically in
\Eqsref{Ndef}]. However, the separability criterion of \cite{GSepCrit}
is now necessary to determine the properties under bipartite
splitting, since for AB-C we deal with a $2\times n$ state and PPT is
then no longer sufficient for biseparability \cite{GBE}.

It is worth pointing out that the separability criterion can be
checked experimentally. The CM $\gamma$ can be measured, and thus the
criterion is entirely formulated in terms of quantities that are
measurable with current technology.

Gaussian CV states promise to be a fruitful testing ground for quantum
nonlocality: Pure entanglement is comparatively easy to create in
quantum optical experiments, as described in \cite{CVghz}. Likewise,
tripartite bound entangled states are experimentally accessible: the
states discussed in the examples Subsec.\ \ref{extmss} and \ref{exghz}
can be obtained by mixing differently displaced pure Gaussian states.

The study of entanglement of multi-party Gaussian states is still in a
very early stage. For example, no work has to our knowledge been done
on the interesting cases of more parties and modes.  But even for the
simple three-mode case there are important open question. In
particular nothing is known about the distillability of tripartite
states. As in Ref. \cite{classent} for qubits, it is easy to see that
Gaussian states in Classes 3 and 4 cannot be distilled at all and are
therefore bound entangled. For this, we consider $N$ copies of a Class
3 state $\rho$, and apply an arbitrary local quantum operation
$\cP_{locc}$ consisting of a classically correlated sequence of
operations of the form $\cP=\cP_A\otimes\cP_B\otimes\cP_C$. Since
$\rho$ is in Class 3 we can write $\rho^{\otimes N}$ as a mixture of
$AB-C$ product states
$\sum_kp_k\rho_{AB,k}^{(N)}\otimes\rho_{C,k}^{(N)}$ and as a mixture
of $AC-B$ product states
$\sum_kp'_k\rho_{AC,k}^{(N)}\otimes\rho_{B,k}^{(N)}$.  After applying
an operation such as $\cP$ the resulting state $\tilde\rho =
\cP(\rho^{\otimes N})$ will still be separable along these cuts, and
no sequence of operations $\cP$ can change this. Thus $\rho$ is bound
entangled.

Whether all states in Class 2 may be distilled to maximally entangled
states between the two non-separable parties is an open question. If
this were shown, it would follow that all states in Class 1 could be
distilled into arbitrary tripartite entangled states.

\begin{acknowledgments}
G.G. acknowledges financial support by the Friedrich-Naumann-Stiftung. B.K.
and J.I.C. thank the University of Hannover for hospitality. M.L., B.K., and
J.I.C. acknowledge the hospitality of the Erwin Schr\"odinger Institute. This
work was supported by the Austrian Science Foundation under the SFB ``Control
and Measurement of Coherent Quantum Systems'' (Project 11), the European
Union under the TMR network ERB--FMRX--CT96--0087 and the project EQUIP
(contract IST-1999-11053), the European Science Foundation, the Institute for
Quantum Information GmbH, Innsbruck, and the Deutsche Forschungsgemeinschaft
(SFB 407 and Schwerpunkt "Quanteninformationsverarbeitung").
\end{acknowledgments}
\appendix

\section{Points of Intersection}\label{app}
As shown in \Thref{fullsepTh} a state is separable iff solutions to
\Ineqsref{sepcond} are found among the points of intersection of the curves
described by the \emph{equalities} (\ref{sepcond}), or the centers of the
three sets. Here we give the formulas to directly calculate these points from
$\gamma$.

The centers of circle and the ellipses have already been
shown to be
\begin{eqnarray}\label{fullsep1}
m_c&=&(0,0)^T,\nonumber\\
m_{e}&=& \frac{\det N
+1}{k_1} L,\\
m_{\tilde e}&=& \frac{\det \tilde N +1}{\tilde k_1} \tilde L,\nonumber
\end{eqnarray}
where $N, \tilde N$ were defined in (\ref{Ndef}), $L$ in (\ref{Ldef}), and
$k_1, \tilde k_1$ after (\ref{ellipse}). The intersections of the borders of
$\cC, \cE, \tilde\cE$ are calculated as follows. Consider first the two
ellipses, whose borders are defined by the \emph{equalities} (\ref{sce},
\ref{scet}). Dividing by $\tr N$, respectively by $\tr\tilde N$ and
subtracting the two equalities we find that a point on both $\del\cE$ and
$\del\tilde\cE$ must lie on the straight line $\cG_{e\tilde e}$ defined by
\begin{equation}\label{gerade}
(\det N+1+L^T\xi)/\tr N = (\det \tilde N+1+\tilde L^T\xi)/\tr
\tilde N,
\end{equation}
where $\xi=(y,z)$.
$\cG_{e\tilde e}$ can be parameterized with $s\in\RR$ as $g_{e\tilde
e}+sf_{e\tilde e}$, where
\begin{equation}
g_{e\tilde e} =
\left( \frac{\det N+1}{\tr N} -\frac{\det \tilde N +1}{\tr\tilde N}
\right)L'/\|L'\|^2,
\end{equation}
where $L'=\tilde L /\tr\tilde N- L/\tr N$ \cite{fnLprime0}
and $f_{e\tilde e}$ is a vector
orthogonal to $L'$.

Inserting $\cG_{e\tilde e}$ in the equation (\ref{sce}) for $\del\cE$
we obtain a quadratic polynomial in $s$, whose roots $s^\pm_{e\tilde
e}$ (if they are real) give the intersection points.
For the intersections of $\del\cC$ with the ellipses we proceed
similarly. In summary, we get for the intersection points
\begin{equation}\label{fullsep2a}
i_{e\tilde e}^{\pm} = g_{e\tilde e}+s_{e\tilde
e}^{\pm} f_{e\tilde e},
\end{equation}
\begin{equation}\label{fullsep2b}
i_{ce}^{\pm} = g_{ce}+s_{ce}^{\pm} f_{ce},
\end{equation}
\begin{equation}\label{fullsep2c}
i_{c\tilde e}^{\pm} = g_{c\tilde e}+s_{c\tilde e}^{\pm} f_{c\tilde e},
\end{equation}
where the vectors $g_x, x=ce, c\tilde e$ are
\begin{subequations}
\begin{eqnarray}
g_{ce} &=& \left(\tr N\sqrt{r_c^2 +1} -\det N -1\right)L/\|L\|^2,
\end{eqnarray}
\end{subequations}
$f_{ce}$ is a vector orthogonal to $L$, and
$r_c$ is the smaller of the two radii
\begin{equation}\label{radius}
r_c = \mathrm{min}\left\{
\sqrt{(\tr N)^2/4-1},\sqrt{(\tr\tilde
N)^2/4-1}\right\}.
\end{equation}
$g_{c\tilde e}, f_{e\tilde e}$
are defined likewise for tilded quantities. And, finally, by
$s^{\pm}_{e\tilde e}, s^\pm_x$
we denote the real roots of the  quadratic polynomials
\begin{subequations}\label{polynomials}
\begin{eqnarray}
P_{e\tilde e}(s) &=& \left( L^T(g_{e\tilde
e}+sf_{e\tilde e}) +\det N +1\right)^2 - \nonumber\\
&&(\tr N)^2\left(
1+\|g_{e\tilde e}+sf_{e\tilde e}\|^2 \right),\\
P_{x}(s) &=& r_c^2-\|g_{x}+sf_{x}\|^2, x=ce, c\tilde e.
\end{eqnarray}
\end{subequations}
Thus all the nine candidates are given in terms of $N, \tilde N$ which
can be directly obtained from $\gamma$.

\section{Characterization of $K$}\label{app2}

Here we show that $K(\gamma)$ as defined in \Eqref{K}
coincides with the (real) span of the vectors belonging to the
kernels of $\gamma+\tilde J_x\gamma^{-1}\tilde J_x$. This fact
automatically follows from the following

\begin{lemma} (Characterization of $K(\gamma)$)\\
 \label{charK}
Let $f=f_R+if_I$, where $f_R$ and $f_I$ are real. Then $f\in
\ker(\gamma-i\tilde J_x)$ iff $f_I=\gamma^{-1} \tilde J_x f_R$ and
both $f_R$ and $f_I$ belong to the kernel of $\gamma+\tilde J_x
\gamma^{-1} \tilde J_x$.
\end{lemma}

\Proof\ Taking the real and imaginary part of the equation
$(\gamma-i\tilde J_x)f=0$ we find $ \gamma f_R +\tilde J_x
f_I=0$ and $\gamma f_I- \tilde J_x f_R=0$. Since $\gamma$ must
be invertible we obtain from the second equation that
$f_I=\gamma^{-1} \tilde J_x f_R$. Using now the first equation
we find that $(\gamma+\tilde J_x\gamma^{-1} \tilde J_x)f_R=0$.
Analogously, $(\gamma+\tilde J_x\gamma^{-1} \tilde J_x)f_I=0$.
The same argumentation holds for the other direction of the
proof.\hfill\qed



\begin{thebibliography}{99}
\bibitem{fndimofJ} To be precise, we should define $J$ with an index $n$ to
keep track of the dimension of the space $\RR^{2n}$ on which it
acts. But since $n$ will always be clear from the context we will omit
this index and just use $J$ to make the expressions more readable.
\bibitem{fnblock}
In the following, it is convenient to use the notation $A\oplus B$ for
block-diagonal matrices: if $A$ and $B$ are $n\times n$ and $m\times m$
square matrices, respectively, then $A\oplus B$ is the $(n+m)\times (n+m)$
square matrix
\[{A\,\,\,0 \choose 0\,\,\, B}.\]
\bibitem{fnLnot0} The following definitions
assume that $L, \tilde L\not=0$. (If one of them is $0$, the corresponding
ellipse degenerates into a circle around $(0,0)$ and we can take an arbitrary
$L\not=0$ to make sense of $P_L$.) The criterion is not affected by this
assumption, since it relies on \Ineqsref{sepcond}.
\bibitem{fnLprime0}
In the case $L'=0$
the borders of the ellipses do either not intersect at all or
coincide. In both cases we have to look for solutions among the
remaining seven candidates.
\end{thebibliography}
\end{document}